\documentstyle[prc,preprint,tighten,aps]{revtex}
\begin {document}
\draft
\title{
Searching for three-nucleon resonances}
\author{A. Cs\'ot\'o$^1$, H. Oberhummer$^2$, and R. Pichler$^2$}
\address{$^1$National Superconducting Cyclotron Laboratory,
Michigan State University, East Lansing, Michigan 48824 \\
$^2$Institut f\"ur Kernphysik, Technische Universit\"at Wien, Wiedner
Hauptstrasse 8--10, 1040 Vienna, Austria}
\date{\today}

\maketitle

\begin{abstract}
\noindent
We search for three-neutron resonances which were predicted from
pion double charge exchange experiments on $^3$He. All partial
waves up to $J={{5}\over{2}}$ are nonresonant except the $J^\pi=
{{3}\over{2}}^+$ one, where we find a state at $E=14$ MeV energy
with 13 MeV width. The parameters of the mirror state in
the three-proton system are $E=15$ MeV and $\Gamma=14$ MeV.
The possible existence of an excited state in the triton, which was
predicted from a H($^6$He,$\alpha$) experiment, is also discussed.
\end{abstract}

\narrowtext

\section{Introduction}

One of the clearest indication of the spin dependence of the
nucleon-nucleon interaction is the fact that there exists a
bound triplet deuteron, but not a singlet deuteron or dineutron.
The possibility that by adding one or two more neutrons to an
unbound dineutron one might wind up with a bound tri-neutron
or tetra-neutron has long been discussed. However, precise
experiments and calculations show no indication for the existence
of such bound structures \cite{Tilley}. Nevertheless, the existence
of resonant states of these nuclei, which have observable effects,
cannot be excluded.

Sperinde {\it et al.} found \cite{Sperinde} that the differential
cross section of the $^3$He($\pi^-,\pi^+)3n$ pion
double charge exchange reaction strongly differs from what one would
expect from a pure phase-space description of the final state.
The discrepancy seemed to be well explained by the assumption that
there exists a three-neutron resonance at 2 MeV energy with 12 MeV
width. The authors suggested that this state is probably a member
of a $T={{3}\over{2}}$ isospin quartet in the $A=3$ nuclei, another
member being the three-proton state found in \cite{Williams}.
The mechanism of the double charge exchange experiment was studied
theoretically in \cite{Jibuti}. The authors found that by taking into
account the final state interaction between the outgoing neutrons,
one can get good agreement with the experimental results of
\cite{Sperinde}. However, the nature of this final state interaction,
whether it produces a three-neutron resonance or not, was not studied in
\cite{Jibuti}.

Recently, a more thorough experimental study of the
$^3$He($\pi^-,\pi^+)3n$ process was performed \cite{Stetz}. The
authors pointed out that the analysis of \cite{Sperinde} was
incorrect, because a four-body phase-space had been divided out of
the cross section, instead of the correct three-body one. And this
mistake leads to a peak in the cross section at low energies.
However, three-neutron structures also appear in \cite{Stetz} at
small angles. These structures at $\sim 20$ MeV three-neutron
missing mass and with $\sim 20$ MeV width could be caused by
three-neutron resonances.

There are experimental indications of the possible existence of
resonances in other $A=3$ nuclei, too. For example, in a recent
H($^6$He,$\alpha$) experiment \cite{Aleksandrov} the authors
concluded that in order to explain their results, an excited state
of the triton has to be assumed with $E^*=7.0\pm 0.3$ MeV excitation
energy and $\Gamma =0.6\pm 0.3$ MeV width.

Theoretically the problem of a three-neutron resonance was studied
by Gl\"ockle \cite{Glockle}, who calculated the pole position of
the $S$ matrix from the analytical continuation of the Faddeev kernel
for three neutrons with pure $^1$$S_0$ $N-N$ interaction. He found
that the pattern of the pole trajectory ruled out the possibility
of a low-energy ${{1}\over{2}}^+$ three-neutron resonance in his model.

The aim of this current work is to systematically investigate the
possible existence of resonances in the lowest partial waves of the
three-neutron system by using the complex scaling method. We shall
also comment on the triton excited state.

\section{Complex scaling}

In coordinate space, resonance eigenfunctions, corresponding to the
complex energy solutions of the Schr\"odinger equation
\begin{equation}
\widehat H\vert\Psi\rangle=(\widehat T+\widehat
V)\vert\Psi\rangle =E\vert\Psi\rangle ,
\label{Sch}
\end{equation}
show oscillatory behavior in the asymptotic region with
exponentially growing amplitude, $\sim \exp [i(\kappa -i\gamma )r]$
$(\kappa ,\gamma >0)$. Thus, the resonance eigenfunctions are not
square-integrable. The complex scaling method (CSM) \cite{CSM}
reduces the description of resonant states to that of bound
states, thus avoiding the problem of asymptotics. In the CSM
we define a new Hamiltonian by
\begin{equation}
\widehat H_\theta=\widehat U(\theta)\widehat H\widehat U^{-1}
(\theta),
\end{equation}
and solve the complex equation
\begin{equation}
\widehat H_\theta\vert\Psi_\theta\rangle=
\varepsilon\vert\Psi_\theta\rangle .
\label{SchCSM}
\end{equation}
Here the unbounded similarity transformation $\widehat U(\theta)$
acts, in the coordinate space, on a function $f({\bf r})$ as
\begin{equation}
\widehat U(\theta)f({\bf r})=e^{3i\theta/2}f({\bf r}e^{i\theta}).
\label{CS}
\end{equation}
(If $\theta$ is real, $\widehat U(\theta)$ means a rotation
into the complex coordinate plane, if it is complex, it means
a rotation and scaling.)
In the case of a many-body Hamiltonian, (\ref{CS}) means that the
transformation has to be performed in each Jacobi coordinate.
For a broad class of potentials there is the following connection
between the spectra of $\widehat H$ and $\widehat H_\theta$ \cite{ABC}:
(i) the bound eigenstates of $\widehat H$ are the
eigenstates of $\widehat H_\theta$, for any
value of $\theta$ within $0\leq\theta<\pi/2$;
(ii) the continuous spectrum of $\widehat H$ will be
rotated by an angle 2$\theta$;
(iii) the complex generalized eigenvalues of
$\widehat H_\theta$, $\varepsilon_{\rm res}=E
-i\Gamma /2$, $E, \Gamma >0$
(where $\Gamma$ is the full width at half maximum)
belong to its proper spectrum, with square-integrable
eigenfunctions, provided $2\theta >\vert \arg
\varepsilon_{\rm res}\vert$. These complex eigenvalues coincide
with the $S$ matrix pole positions. This method was tested
for three-body resonances in \cite{3b}, and was applied to
the low-lying spectra of the $^6$He, $^6$Li, and $^6$Be
nuclei in \cite{soft}. Further details and references of
the method can be found there.

Up to Eq.\ (\ref{SchCSM}) our treatment of the three-body
resonances is exact. The only approximation we use here is
that now we expand the wave function of Eq.\ (\ref{SchCSM})
in terms of products of Gaussian functions with different widths.
Thus, we select the square integrable solutions of Eq.\ (\ref{SchCSM}),
which are the three-body resonances, and discretize the continuum.
A term of this expansion looks like $\rho_1^{l_1}
\exp [-(\rho_1/\gamma_i)^2]Y_{l_1m_1}(\widehat
\rho_1)\cdot \rho_2^{l_2}
\exp [-(\rho_2/\gamma_j)^2]Y_{l_2m_2}
(\widehat \rho_2)$,
where $l_1$ and $l_2$ are the angular momenta in the
two relative motions, respectively, and the widths
$\gamma$ of the Gaussians are the parameters of the expansion.

\section{Results}

We use the Minnesota effective nucleon-nucleon interaction \cite{MN},
together with the tensor force of \cite{tens,PRL}. The space exchange
parameter of the tensor force is adjusted to the splitting of the
$^3$$P_J$ phase shifts. As one can see in Fig.\ 1, this force produces
$p+p$ phase shifts which are in good agreement with experiment. The
higher partial waves, not shown in Fig.\ 1, are also close to the
(practically zero) experimental values. In the three-neutron wave
functions all $n+n$ partial waves up to $l=2$ are included. In some
cases we checked the role of the $l=3$ partial waves and found
them insignificant.

Figure 2(a) shows a typical result of our calculations, in this
case for $J^\pi={{3}\over{2}}^-$. What one can see is the complex
eigenvalue spectra of Eq.\ (\ref{SchCSM}), with
$E={\rm Re}(\varepsilon)$ and $\Gamma
=2{\rm Im}(\varepsilon)$. The eigenvalues are all sitting
on a half-life whose angle to the $E$ axis is roughly $2\theta$. This
line is the discretized rotated continuum of Eq.\ (\ref{SchCSM}).
If there were a resonance in this partial wave, it would be in
the upper right triangle, separated from the continuum points.

In Fig.\ 2(b) we show the result for $J^\pi={{3}\over{2}}^+$, which
is found to be the only resonant partial wave. The resonant state is
shown by the circle. We find that the dominant terms in the
resonant $J^\pi={{3}\over{2}}^+$ wave function are the
$[(l_1l_2)L,S]=[(11)1,{{3}\over{2}}]$ and $[(11)2,{{1}\over{2}}]$
$LS$ components, where $L$ and $S$ are the the total angular momentum
and total spin, respectively. The other components have small but
non-negligible effects.

Because we solve Eq.\ (\ref{SchCSM}) approximately, the resonance
papameters depend slightly on $\theta$. To find the optimal $\theta$
value one should repeat the calculations with several different
$\theta$ values, and find a region where the resonance parameters are
most insensitive to the change of $\theta$. This way of finding the
optimal $\theta$ value is related to the complex virial theorem
\cite{virial}. To reduce computer time, we perform these
calculations with a wave function which contains only the above
mentioned two most important $LS$ components. In Fig.\ 3 one can see
that the optimal $\theta$ angle is somewhere between 0.3 and 0.35 rad.
Using such a $\theta$ value in the full model the parameters of the
resonance are $E=14$ MeV and $\Gamma=13$ MeV.

What can one say about possible resonances in other $A=3$ nuclei?
The effect of the Coulomb force in the mirror three-proton system
is trivial, pushing the ${{3}\over{2}}^+$ resonance toward higher
energy and increasing its width accordingly. The resonance parameters
are $E=15$ MeV and $\Gamma=14$ MeV.

The situation is more complex
in the cases of $^3$H and $^3$He, because these nuclei have two-body
channels below the three-nucleon thresholds. Recently an experiment,
studying the H$($$^6$He,$\alpha$) reaction, found indications of an
excited state of the triton at $E^*=7.0\pm 0.3$ MeV excitation energy
with $\Gamma =0.6\pm 0.3$ MeV width \cite{Aleksandrov}. This state
would be situated between the $d+n$ and $n+n+p$ thresholds. It
was suggested in \cite{Aleksandrov} and \cite{Barabanov} that the
dominant configuration in this state is the $(nn)p$ one, i.e., the
resonance originates from the closed three-body channel.

Unfortunately our current $N-N$ interaction is not appropriate for
a calculation which contains all the relevant angular momentum
configurations for the triton. This is because the Minnesota
force reproduces the $^3$$S_1$ $n+p$ phase shift and the deuteron
binding energy without the tensor force. However, we can estimate
the role of the $n+n+p$ and $d+n$ channels in a restricted model,
which contains $^3$$D_1$ states just between the $d$ and $n$ but
not inside the deuteron. Th Minnesota force is shown \cite{Tang}
to give a good overall description for the $d+p$, and supposedly
for the $d+n$, scattering in such a model. It is important to
note that, if there is a resonance in the $(nn)p$ configuration,
it must appear in a one-channel $d+n$ model, too. The reason is
that any $J^\pi$ state that can be formed in the $(nn)p$
configuration can be built up from $d+n$ one, too, and this $d+n$
channel is nonorthogonal to the $(nn)p$ configuration. Thus, as
the CSM is not well suited for finding poles at energies with
negative real and imaginary parts (the $E^*=7.0$ MeV energy would mean
such a position in the $n+n+p$ channel), we look for poles in the
$d+n$ scattering. We use the analytic continuation of the $S$ matrix
to complex energies \cite{PRL}, which is feasible for a two-body
case and equivalent with the CSM. A ${{1}\over{2}}^+$ pole is found at
$E=1.4$ MeV $d+n$ center-of-mass energy with a width of 9 MeV. The
$n+n+p$ channel can be taken into account approximately, by using a
few discretized continuum state of $(nn)$ with positive energies
and square-integrable wave functions. The inclusion of this $n+n+p$
channel does not change the pole position significantly. The dominant
configuration of the wave function is $d+n$, in contrast to the
case of \cite{Barabanov}.

The quality of our $^3$$S_1$ and $^3$$D_1$ $N-N$ forces makes it
impossible too reach a firm conclusion concerning the existence
of a narrow excited state of the triton. Our main result is that
the picture of \cite{Aleksandrov} and \cite{Barabanov}, namely
that this state originates from the $(nn)p$ channel, is
questionable. As such a state would be nonorthogonal to the $d+n$
channel, this latter configuration must be important.

Finally, we note that the use of realistic interactions together
with a full model space (including $l\ge 3$) may change the
parameters of the ${{3}\over{2}}^+$ resonance. This is why we have
made no attempt to further optimize our basis (expansion length,
Gaussian widths, etc.) and find more precise resonance parameters.
Our aim was to show the existence of such a state.

\section{Conclusion}

In summary, we have searched for three-neutron resonances using
the complex scaling method. We have found a $J^\pi ={{3}\over{2}}^+$
resonant state at $E=14$ MeV energy with $\Gamma =13$ MeV width.
All other partial waves up to $J={{5}\over{2}}$ were found to be
nonresonant. The parameters of the mirror ${{3}\over{2}}^+$
three-proton resonance are $E=15$ MeV and $\Gamma =14$ MeV. Our
results are inconclusive regarding the narrow excited state of
the triton, but show the importance of the $d+n$ channel.

\acknowledgments

The idea of this work came from discussions with Roman Kezerashvili
and Scott Pratt. The work of A.\ C.\ was supported by Wolfgang Bauer's
Presidential Faculty Fellowship (PHY92-53505) and NSF Grant No.\
PHY94-03666. We also want to thank the Fonds zur F\"orderung
wissenschaflicher Forschung in \"Osterreich (project P 10361--PHY)
for their support.

\narrowtext
\begin{figure}
\caption{$S$-wave (a) and $P$- and $D$-wave (b) $p+p$ phase shifts
in the center-of-mass frame. Experimental data are taken from
\protect\cite{Arndt}.}
\label{fig1}
\end{figure}

\narrowtext
\begin{figure}
\caption{Energy eigenvalues of the complex scaled Hamiltonian
of the (a) ${{3}\over{2}}^-$, and (b) ${{3}\over{2}}^+$
three-neutron states; $E={\rm Re}(\varepsilon)$,
$\Gamma=2\;{\rm Im}(\varepsilon)$. The dots are the points of the
rotated discretized continua, while the circle is a
three-neutron resonance. The rotation angle is 0.4 rad. }
\label{fig2}
\end{figure}

\narrowtext
\begin{figure}
\caption{The $\theta$ trajectory of the ${{3}\over{2}}^+$
three-neutron resonance with a two-component wave function.
The $\theta$ values are indicated.}
\label{fig3}
\end{figure}


\begin{references}
\bibitem{Tilley} D.~R. Tilley, H.~R. Weller, and H.~H. Hasan,
Nucl. Phys. {\bf A474}, 1 (1987).
\bibitem{Sperinde} J. Sperinde, D. Frederickson, R. Hinkins,
V. Perez-Mendez, and B. Smith, Phys. Lett. {\bf 32B}, 185 (1970).
\bibitem{Williams} L.~E. Williams, C.~J. Batty, B.~E. Bonner,
C.~Tschal\"ar, H.~C. Ben\"ohr, and A.~S. Clough, Phys. Rev. Lett.
{\bf 23}, 1181 (1969).
\bibitem{Jibuti} R.~I. Jibuti and R. Ya. Kezerashvili,
Nucl. Phys. {\bf A437}, 687 (1985).
\bibitem{Stetz} A. Stetz {\it et al.}, Nucl. Phys. {\bf A457},
669 (1986).
\bibitem{Aleksandrov} D.~V. Aleksandrov, E.~Yu. Nikol'skii,
B.~G. Novatskii, and D.~N. Stepanov, JETP Lett. {\bf 59}, 320
(1994).
\bibitem{Glockle} W. Gl\"ockle, Phys. Rev. C {\bf 18}, 564 (1978).
\bibitem{CSM} Y.~K. Ho, Phys. Rep. {\bf 99}, 1 (1983);
N. Moiseyev, P.~R. Certain, and F. Weinhold,
Mol. Phys. {\bf 36}, 1613 (1978); Proceedings of the Sanibel
Workshop Complex Scaling, 1978 [Int. J. Quantum Chem. {\bf 14},
343 (1978)]; B.~R. Junker, Adv. At. Mol. Phys. {\bf 18}, 207
(1982); W.~P. Reinhardt, Annu. Rev. Phys. Chem. {\bf
33}, 223 (1982); {\em Resonances--The Unifying Route Towards
the Formulation of Dynamical Processes, Foundations and
Applications in Nuclear, Atomic and Molecular Physics},
edited by E. Br\"andas and N. Elander, Lecture Notes in Physics
Vol. 325 (Springer-Verlag, Berlin, 1989).
\bibitem{ABC} J. Aguilar and J.~M. Combes, Commun. Math. Phys.
{\bf 22}, 269 (1971); E. Balslev and J.~M. Combes, {\em ibid.}
{\bf 22}, 280 (1971); B. Simon, {\em ibid.} {\bf 27}, 1 (1972).
\bibitem{3b} A. Cs\'ot\'o, Phys. Rev. C {\bf 49}, 2244 (1994).
\bibitem{soft} A. Cs\'ot\'o, Phys. Rev. C {\bf 49}, 3035 (1994).
\bibitem{MN} D.~R. Thompson, M. LeMere and Y.~C. Tang, Nucl.
Phys. {\bf A268}, 53 (1977); I. Reichstein and Y. C. Tang,
Nucl. Phys. {\bf A158}, 529 (1970); A. Cs\'ot\'o, Phys. Rev.
C {\bf 48}, 165 (1993).
\bibitem{tens} P. Heiss and H.~H. Hackenbroich, Phys. Lett.
{\bf 30B}, 373 (1969).
\bibitem{PRL} A. Cs\'ot\'o, R.~G. Lovas, and
A.~T. Kruppa, Phys. Rev. Lett. {\bf 70}, 1389 (1993).
\bibitem{Arndt} R.~A. Arndt, R.~H. Hackman, and L.~D.
Roper, Phys. Rev. C {\bf 15}, 1002 (1977)
\bibitem{virial} E. Br\"andas and P. Froelich, Phys. Rev. A
{\bf 16}, 2207 (1977); R. Yaris and P. Winkler, J. Phys.
{\bf B11}, 1475 (1978).
\bibitem{Barabanov} A.~L. Barabanov, JETP Lett. {\bf 61}, 7
(1995).
\bibitem{Tang} T. Kaneko, H. Kanada, and Y.~C. Tang, Few-body
Syst. {\bf 18}, 1 (1995).
\end{references}
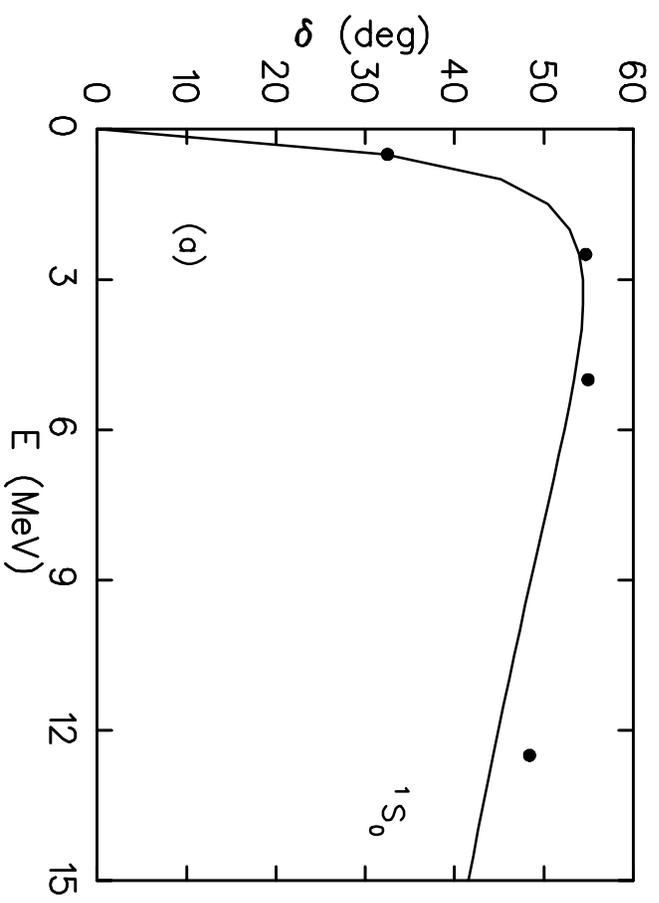
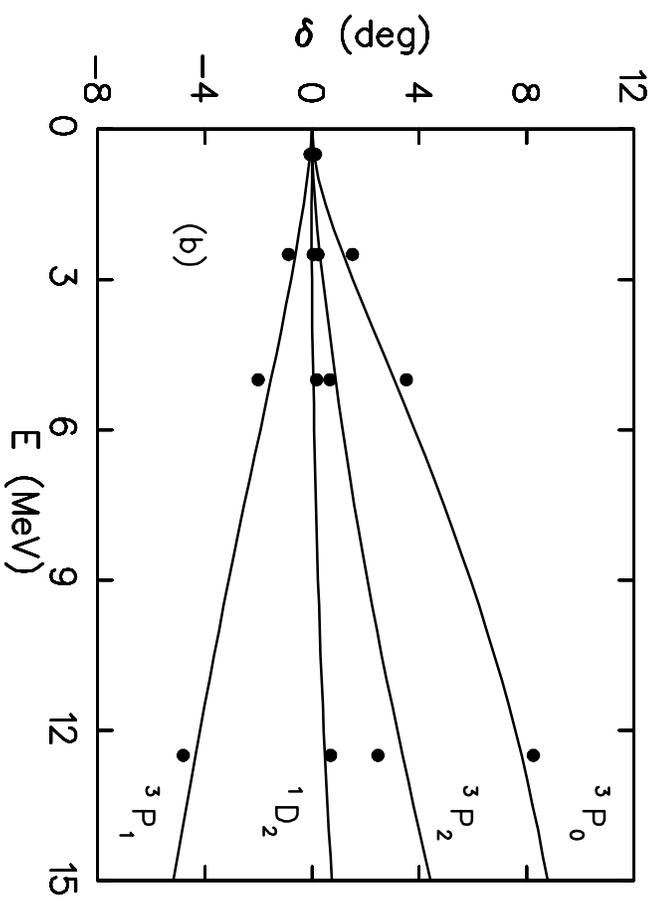
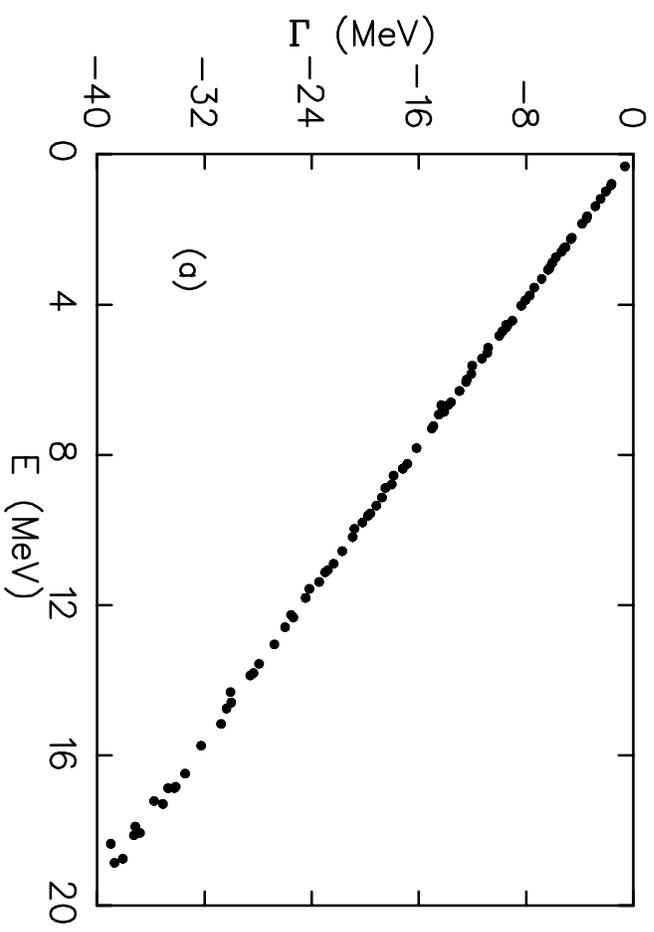
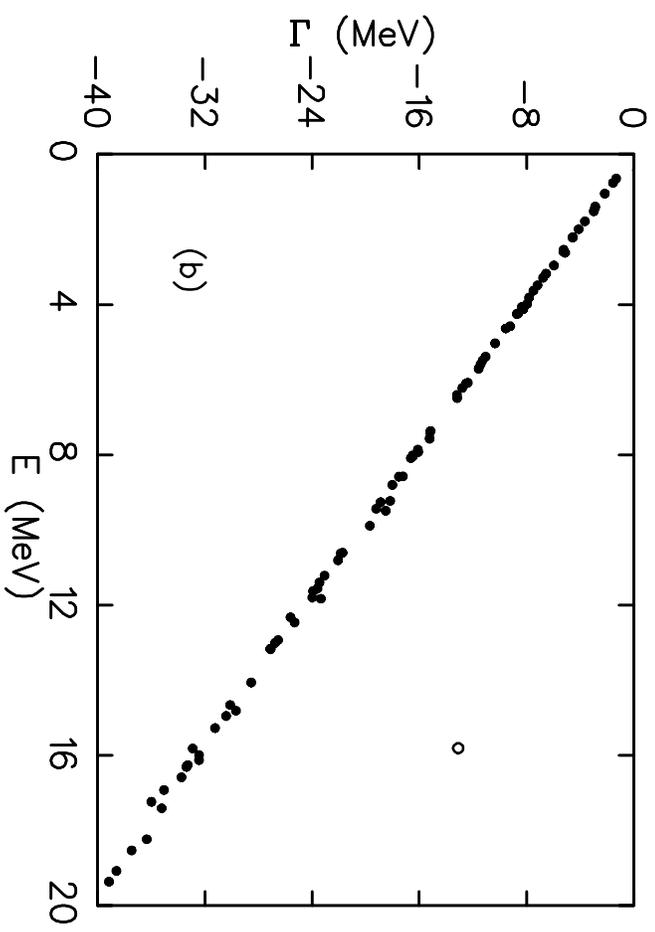
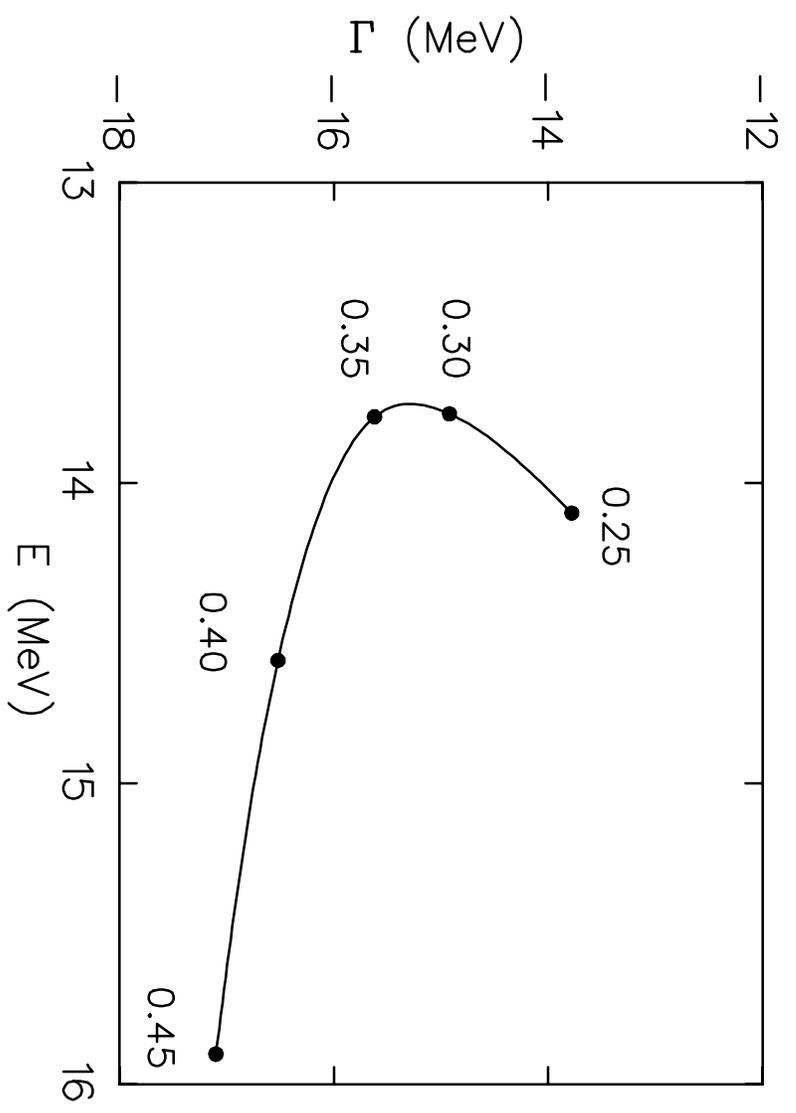
\end{document}